\documentclass[a4paper,11pt]{article}

\usepackage{amsmath,amssymb}     
\usepackage{color}
\usepackage{graphicx}
\usepackage{subfigure}
\usepackage{cite}                
\usepackage{hyperref}            
\usepackage{multirow,makecell}   

\numberwithin{equation}{section}   

\def \be {\begin{equation}}
\def \ee {\end{equation}}
\def \ba {\begin{array}}
\def \ea {\end{array}}
\def \bea{\begin{eqnarray}}
\def \eea{\end{eqnarray}}
\def \nn {\nonumber}

\def \a {\alpha}
\def \b {\beta}
\def \g {\gamma}

\def \G {\Gamma}
\def \d {\delta}

\def \e {\epsilon}

\def \s {\sigma}
\def \S {\Sigma}
\def \r {\rho}

\def \th {\theta}

\def \mA {\mathcal A}
\def \mB {\mathcal B}
\def \mC {\mathcal C}
\def \mD {\mathcal D}
\def \mE {\mathcal E}
\def \mF {\mathcal F}

\def \mH {\mathcal H}
\def \mI {\mathcal I}

\def \mN {\mathcal N}

\def \mP {\mathcal P}

\def \mR {\mathcal R}

\def \p {\partial}
\def \f {\frac}

\def \mc {\mathcal}

\def \lt {\left}
\def \rt {\right}

\def \td {\tilde}

\def \inf {\infty}

\def \lag {\langle}
\def \rag {\rangle}
\def \ph  {\phantom}

\def \ep {\mathrm{e}}
\def \ii {\mathrm{i}}

\def \tr {\textrm{tr}}

\def \and {{\textrm{and}}}

\def \vac {{\langle0|}}
\def \uum {{|0\rangle}}

\def \cl {{\textrm{cl}}}
\def \tree {{\textrm{tree}}}
\def \oloop {{\textrm{1-loop}}}
\def \tloop {{\textrm{2-loop}}}
\def \Area {{\textrm{Area}}}
\def \NS {{\textrm{NS}}}
\def \R {{\textrm{R}}}

\begin{document}

\title{\textbf{Holographic R\'enyi entropy for two-dimensional $\bf{\mathcal N=(1,1)}$ superconformal field theory}}
\author{Jia-ju Zhang\footnote{jjzhang@ihep.ac.cn}}
\date{}

\maketitle

\vspace{-10mm}

\begin{center}
{\it
 Theoretical Physics Division, Institute of High Energy Physics, Chinese Academy of Sciences,\\
19B Yuquan Rd, Beijing 100049, P.~R.~China\\ \vspace{1mm}
Theoretical Physics Center for Science Facilities, Chinese Academy of Sciences,\\19B Yuquan Rd, Beijing 100049, P.~R.~China
}
\vspace{10mm}
\end{center}

\begin{abstract}

  In this paper we investigate the holographic R\'enyi entropy in $\mathcal N=1$ supergravity (SUGRA) in AdS$_3$ spacetime, which is dual to the two-dimensional $\mathcal N=(1,1)$ superconformal field theory (SCFT). We consider both cases of two short intervals on a line with zero temperature and one interval on a circle with low temperature. In SUGRA side we consider contributions of both graviton and gravitino, and in SCFT side we consider contributions of both stress tensor $T$, $\bar T$ and their superpartners $G$, $\bar G$. We find matches between SUGRA and SCFT results.

\end{abstract}

\baselineskip 18pt
\thispagestyle{empty}
\newpage

\tableofcontents

\section{Introduction}

The investigation of entanglement entropy has been gaining more and more attention in the last decade.
We give the definitions of entanglement entropy \cite{nielsen2010quantum,petz2008quantum}.
For a system with normalized density matrix $\r$ with $\tr \r=1$, one can divide the system into a subsystem $A$ and its complement $B$. The entanglement is defined as
\be
S_A=-\tr_A \r_A \log\r_A,
\ee
with the reduced density matrix being $\r_A=\tr_B\r$. It encodes the quantum entanglement between $A$ and $B$.
To calculate the entanglement entropy, one can use the replica trick \cite{Callan:1994py}. One firstly calculates the R\'enyi entropy
\be
S_A^{(n)}=-\f{1}{n-1}\log\tr_A\r_A^n,
\ee
and then takes $n\to1$ limit to get the entanglement entropy. For two subsystems $A$ and $B$ that are not necessarily complements of each other, one can define the mutual information
\be
I_{A,B}=S_A+S_A-S_{A\cup B},
\ee
and the R\'enyi mutual entropy
\be
I_{A,B}^{(n)}=S_{A}^{(n)}+S_{B}^{(n)}-S_{A\cup B}^{(n)}.
\ee
When there is no ambiguity, we write for short $S=S_A$, $S_n=S_A^{(n)}$, $I=I_{A,B}$, and $I_n=I_{A,B}^{(n)}$.

The R\'enyi entropies in a two-dimensional conformal field theory (CFT) are much easier to calculate than their higher dimensional cousins. For the case of one interval on an infinite straight line with zero temperature, the R\'enyi entropy is exact and universal \cite{Calabrese:2004eu}
\be
S_n=\f{c(n+1)}{6n}\log\f{l}{\e},
\ee
with $c$ being central charge of the CFT, $l$ being the length of the interval, and $\e$ being the UV cutoff. For cases of multiple intervals, there are no universal results and details of the CFT are needed \cite{Caraglio:2008pk,Furukawa:2008,Calabrese:2009ez,Headrick:2010zt,Calabrese:2010he}. However, when the interval is short perturbative calculation is available \cite{Headrick:2010zt,Calabrese:2010he,Chen:2013kpa}. Similarly, R\'enyi entropy of one interval on a circle with zero temperature is universal
\cite{Calabrese:2004eu}
\be\label{e12}
S_n=\f{c(n+1)}{6n}\log \Big( \f{R}{\pi\e}\sin\f{\pi l}{R} \Big),
\ee
with $l$ and $R$ being lengthes of the interval and the circle, respectively. When low temperature is turned on, there would be thermal corrections to the R\'enyi entropy that depend on field contents of the CFT \cite{Herzog:2012bw,Herzog:2013py,Cardy:2014jwa,Chen:2014unl,Chen:2015uia}.

As an application of the AdS/CFT correspondence \cite{Maldacena:1997re,Gubser:1998bc,Witten:1998qj,Aharony:1999ti}, one can calculate the entanglement entropy in a CFT using the holographic entanglement entropy \cite{Ryu:2006bv,Ryu:2006ef,Nishioka:2009un,Takayanagi:2012kg} in anti-de Sitter (AdS) spacetime. For a subsystem $A$ in the boundary of the AdS spacetime, the holographic entanglement entropy is proportional to area of the minimal surface $\S_A$ in the bulk that is homogenous to $A$
\be
S_A=\f{\Area(\S_A)}{4G},
\ee
with $G$ being the Newton constant. The area law of the Ryu-Takayanagi (RT) formula for the holographic entanglement entropy has been proved in terms of the generalized gravitational entropy \cite{Lewkowycz:2013nqa}. The RT formula is proportional to the inverse of Newton constant and so is only the classical result, and there are also subleading quantum corrections \cite{Headrick:2010zt,Barrella:2013wja,Faulkner:2013ana}.

It was proposed long time ago that quantum gravity in AdS$_3$ spacetime is dual to a two-dimensional CFT with central charge \cite{Brown:1986nw}
\be
c=\f{3\ell}{2G},
\ee
where $\ell$ is the AdS radius. Expansion of small Newton constant in gravity side corresponds to expansion of large central charge in CFT side. The RT formula in AdS$_3$/CFT$_2$ correspondence was analyzed carefully in both the CFT and gravity sides.
In CFT side the tree level R\'enyi entropy at large central charge is related to Virasoro vacuum block, which only depends on operators in conformal family of identity operator \cite{Hartman:2013mia}. The corresponding classical gravitational configuration was constructed in \cite{Faulkner:2013yia}. Furthermore, 1-loop corrections to holographic R\'enyi entropy under this gravitational background were calculated in \cite{Barrella:2013wja}, and the field contents of gravity theory are relevant.

Various conditions have been investigated for the holographic entropy of two short intervals on a line with zero temperature \cite{Headrick:2010zt,Barrella:2013wja,Chen:2013kpa,Chen:2013dxa,Perlmutter:2013paa,Chen:2014kja,Beccaria:2014lqa,Headrick:2015gba}. One could get the R\'enyi entropy in expansion of the small cross ration $x$. There were calculations in both the gravity and CFT sides, and perfect matches were found.
In case of the correspondence between pure Einstein gravity and a large central charge CFT, there are contributions of graviton in gravity side and contributions of stress tensor $T$, $\bar T$ in CFT side \cite{Headrick:2010zt,Barrella:2013wja,Chen:2013kpa,Chen:2013dxa,Headrick:2015gba}.
In case of higher spin gravity/CFT with $W$ symmetry correspondence, with higher spin chemical potential being turned off, there are contributions of graviton and higher spin fields in gravity side and contributions of stress tensor $T$, $\bar T$ and $W$, $\bar W$ operators in CFT side \cite{Chen:2013dxa,Perlmutter:2013paa,Beccaria:2014lqa,Headrick:2015gba}.
In case of critical massive gravity/logarithmic CFT correspondence, there are contributions of graviton and logarithmic modes in gravity side and contributions of stress tensor $T$, $\bar T$ and their logarithmic partners in CFT side \cite{Chen:2014kja}.
A scalar field in the bulk corresponds to a scalar operator in the CFT, and the 1-loop holographic entanglement entropy of the case was considered in \cite{Beccaria:2014lqa}.
There is a similar story for the holographic R\'enyi entropy of one interval on a circle with low temperature \cite{Barrella:2013wja,Cardy:2014jwa,Chen:2014unl,Chen:2015uia}.

In this paper we extend the previous results to supersymmetric AdS$_3$/CFT$_2$ correspondence. In gravity side we consider $\mN=1$ supergravity (SUGRA) in AdS$_3$ spacetime, where there is massless graviton as well massless gravitino. The $\mN=1$ SUGRA is dual to a two-dimensional $\mN=(1,1)$ superconformal field theory (SCFT), where we have to consider the stress tensor $T$, $\bar T$ and their superpartners $G$, $\bar G$.
We calculate the R\'enyi entropy of both cases of two short intervals on a line with zero temperature and one interval on a circle with low temperature in both SUGRA and SCFT sides. For the case of two intervals, we get the R\'enyi mutual information to order $x^5$ with $x$ being the cross ratio. For the case of one interval on a circle, we get the R\'enyi entropy to order $\ep^{-5\pi\b/R}$ with $\b$ being the inverse temperature and $R$ being the length of the circle. There are perfect matches between SUGRA and SCFT results for both cases.

When dealing with a two-dimensional SCFT, one should be careful with the boundary conditions of fermionic operators.
For an SCFT on a cylinder, one can consider antiperiodic boundary condition of fermionic operators, and it is called Neveu-Schwarz (NS) sector of the SCFT. Or one can consider periodic boundary condition, and it is called Ramond (R) sector.
One can use a conformal transformation and map an SCFT on a cylinder to an SCFT on a complex plane. For an SCFT on a complex plane, in NS sector fermionic operators are periodic when circling around the origin, while in R sector fermionic operators are antiperiodic. Fermionic operators are expanded by half-integer modes in NS sector, and by integer modes in R sector.
In NS sector vacuum of an SCFT on a complex plane one has the conformal weights $h_{\NS}=\bar h_{\NS}=0$, and in R sector vacuum one has $h_\R=\bar h_{\R}=\f{c}{24}$ with $c$ being the central charge. In NS vacuum of an SCFT on a cylinder one has the energy $H_\NS=-\f{\pi c}{6R}$ with $R$ being the circumference of the cylinder, and in R sector vacuum one has $H_\R=0$. In large central charge limit, R sector vacuum is highly excited compared to NS vacuum, and so contributions of R sector to R\'enyi entropy are highly repressed.
In AdS$_3$/CFT$_2$ correspondence, NS sector SCFT corresponds to quantum gravity in global AdS$_3$ spacetime, while on the other hand R sector SCFT corresponds to quantum gravity in background of zero mass BTZ black hole \cite{Coussaert:1993jp,Maldacena:1998bw}. Thus the gravitational configuration in \cite{Faulkner:2013yia,Barrella:2013wja} only corresponds to the NS sector SCFT. So in this paper we only consider contributions of NS sector to R\'enyi entropy.

The rest of the paper is arranged as follows.
In Section~\ref{s2} we give some basic properties of the two-dimensional $\mN=(1,1)$ SCFT.
In Section~\ref{s3} we calculate the R\'enyi mutual information for the case of two intervals on a line with zero temperature in both the SUGRA and SCFT sides.
In Section~\ref{s4} we consider the case of one interval on a circle with low temperature.
We conclude with discussion in Section~\ref{s5}.
We collect some useful summation formulas in Appendix~\ref{sa}.

\section{Basics of two-dimensional $\mN$=(1,1) SCFT}\label{s2}

In this section we give some basic properties of the two-dimensional $\mN=(1,1)$ SCFT on a complex plane that are useful in this paper. One can see details in, for example, the textbooks \cite{DiFrancesco:1997nk,Polchinski:1998rr,Blumenhagen:2009zz}.

In the two-dimensional $\mN=(1,1)$ SCFT, one has the stress tensor $T(z)$ and $\bar T(\bar z)$ and their superpartners $G(z)$ and $\bar G(\bar z)$. Operator $G(z)$ is a holomorphic primary operator with conformal weights $h=3/2$, $\bar h=0$, and $\bar G(\bar z)$ is an antiholomorphic primary operator with conformal weights $h=0$, $\bar h=3/2$. Since the holomorphic and antiholomorphic parts are independent and similar, we will only discuss the holomorphic part below.

For holomorphic quasiprimary operators $\phi_i$ we have two-point function
\be
\lag \phi_i(z)\phi_j(w) \rag_C=\f{\a_i \d_{ij}}{(z-w)^{2h_i}},
\ee
with $C$ denoting the complex plane, $h_i$ being conformal weight of $\phi_i$, and $\a_i$ being the normalization factor.
Note that the operators $\phi_i$ are orthogonalized but not normalized.
Of course we have $\a_1=1$ for the identity operator 1.
For a holomorphic quasiprimary operator $\phi$ with conformal weight $h$ and normalization factor $\a_\phi$, we have the mode expansion
\be
\phi(z)=\sum_r \f{\phi_r}{z^{r+h}},
\ee
with $r$ being integers or half-integers when $h$ is an integer or a half-integer. Note that we only consider NS sector for fermionic operators in this paper. When $\phi$ is hermitian, we have $\phi_r^\dagger=\phi_{-r}$. We also have
\be
\phi_r \uum=0, ~~ r>-h,
\ee
with $\uum$ being the vacuum state. We have the correspondence between operators and states
\be
\p^m\phi \leftrightarrow
|\p^m\phi\rag \equiv \p^m\phi(0)\uum= m!\phi_{-h-m}\uum, ~~ m=0,1,2,\cdots.
\ee
We have the bra states
\be
\lag \p^m\phi| = |\p^m\phi\rag ^\dagger =m!\vac\phi_{h+m}= \vac \p^m\phi(\inf),
\ee
with the definitions
\be \label{pmphiinf}
\p^m\phi(\inf) \equiv \lim_{z\to\inf}(-z^2\p_z)^m [z^{2h}\phi(z)].
\ee
We have the normalization factors
\be \label{apmphi}
\a_{\p^m\phi}\equiv \lag \p^m\phi |\p^m\phi\rag=\f{m!(2h+m-1)!}{(2h-1)!}\a_\phi.
\ee
For examples, when $m=0,1,2$ we have
\bea
&& \phi(\inf) = z^{2h}\phi(z), ~~~ \p\phi(\inf) =  -z^{2h+2}\p\phi(z)-2h z^{2h+1}\phi(z),             \\
&& \p^2\phi(\inf) =  z^{2h+4}\p^2\phi(z)  +2(2h+1) z^{2h+3}\p\phi(z)  +2h(2h+1) z^{2h+2}\phi(z),      \nn
\eea
with the limit $z \to \inf$. Note that the products of bra and ket states can be written as correlation functions. For example we have
\be
\lag \p^m\phi_i |\p^n\phi_j\rag= \lag \p^m\phi_i(\inf) \p^n\phi_j(0) \rag_C=\a_{\p^m\phi_i}\d_{ij}\d^{mn}.
\ee
This strategy has been used in \cite{Chen:2014unl,Headrick:2015gba,Chen:2015uia} to calculate correlation functions.

For quasiprimary operator $T$ and primary operator $G$, we adopt the usual normalization factors
\be
\a_T=\f{c}{2}, ~~~
\a_G=\f{2c}{3},
\ee
with $c$ being the central charge of the SCFT.
As stated in the introduction, we only consider NS sector of the SCFT, and so we expand $G$ by half-integer modes.
We use $L_m$ with $m \in Z$ and $G_r$ with $r \in Z+1/2$ to denote the modes of the $T(z)$ and $G(z)$, and then we have the $\mN=1$ super Virasoro algebra
\bea
&& [L_m,L_n]=(m-n)L_{m+n}+\f{c}{12}m(m^2-1)\d_{m+n},                       \nn\\
&& [L_m,G_r]=\Big( \f{m}{2}-r \Big) G_{m+r},                             \\
&& \{G_r,G_s\}=2L_{r+s}+\f{c}{3}\Big( r^2-\f{1}{4} \Big)\d_{r+s}.        \nn
\eea

Note that every local operator in a two-dimensional unitary CFT can be written as linear combinations of quasiprimary operators and their derivatives.\footnote{For the $\mN=(1,1)$ SCFT, we can also introduce a complex Grassmann variable $\th$ and work in a superspace with coordinate $(z,\th)$. The quasiprimary operators can be combined as super quasiprimary operators in superspace. Each holomorphic super quasiprimary operator is composed of two holomorphic quasiprimary operators $\Phi(z,\th)=\phi(z)+\th\psi(z)$, and they are related by $|\psi\rag \sim G_{-1/2} |\phi\rag$. This may be useful in the search of quasiprimary operators in higer levels. Also for the SCFT$^n$ that will be introduced in Subsection~\ref{s3.2}, we may expand the twistor operators in global superconformal blocks instead of the ordinary global conformal blocks. This may be more convenient in the expansion to higher orders. We thank the anonymous referee for suggestion about this.}
We count the number of linearly independent holomorphic operators in the $\mN=(1,1)$ SCFT to level 5 as
\be\label{trxl0}
\tr x^{L_0}=\prod_{m=0}^\inf\f{1+x^{m+3/2}}{1-x^{m+2}}= 1+x^{3/2}+x^2+x^{5/2}+x^3+2 x^{7/2}+3 x^4+3 x^{9/2}+3 x^5+O(x^{11/2}),
\ee
from which we get the number of holomorphic quasiprimary operators to level 5 as
\be
(1-x)\tr x^{L_0}+x=1+x^{3/2}+x^2+x^{7/2}+2 x^4+x^{9/2}+O(x^{11/2}).
\ee
These operators can be written as quasiprimary operators or derivatives thereof, and they are listed in Table~\ref{counting}.
\begin{table}
\centering
\begin{tabular}{|c|c|c|c|c|c|c|c|c|c|c|}
\hline
level    & 0 & 3/2   & 2   & 5/2    & 3      & 7/2            & 4                     & 9/2             & 5  & $\cdots$   \\\hline
operator & 1 & $G$   & $T$ & $\p G$ & $\p T$ & $\mB$, $\p^2G$ & $\mA$, $\mC$, $\p^2T$ & $\mD$, $\p\mB$, $\p^3G$ & $\p\mA$, $\p\mC$, $\p^3T$ & $\cdots$\\\hline
\end{tabular} \caption{The linearly independent holomorphic operators in $\mN=(1,1)$ SCFT.}\label{counting}
\end{table}

Here there are definitions of quasiprimary operators
\bea\label{e1}
&& \mA=(TT)-\f{3}{10}\p^2T, ~~~ \mB=(TG)-\f{3}{8}\p^2G,          \nn\\
&& \mC=(G\p G)+\f{1}{2(5c+22)} [ 34(TT)-(7c+41)\p^2 T ],         \\
&& \mD=(T\ii\p G)-\f{3}{4}(\ii\p T G)-\f{1}{5}\ii\p^3G,          \nn
\eea
with the brackets denoting normal ordering. Note that $\mC$ is not only a quasiprimary operator, but also a primary one.
The normalization factors for these quasiprimary operators are
\bea
&& 
   \a_\mA=\frac{c (5c + 22)}{10}, ~~~
   \a_\mB=\frac{c (4 c+21)}{12}, \\
&& \a_\mC=\frac{c (4 c+21) (10 c-7)}{6 (5 c+22)}, ~~~
   \a_\mD=\frac{7 c (10 c-7)}{40},                      \nn
\eea
and the normalization factors for the derivatives of the quasiprimary operators can be got easily from (\ref{apmphi}).
In (\ref{e1}) we have added a factor $\ii$ in the definition of $\mD$, and this makes that
\be
\lag \mD(z)\mD(w) \rag_C=\f{\a_\mD}{(z-w)^9},
\ee
with $\a_\mD$ being positive in large $c$ limit. For the same reason we do not have factor $\ii$ in the definition of $\mC$.

Also we need how these operators transform under a general conformal transformation $z \to f(z)$.
We have the Schwarz derivative
\be
s(z) \equiv \f{f'''(z)}{f'(z)}-\f{3}{2} \lt( \f{f''(z)}{f'(z)} \rt)^2,
\ee
and define the shorthand
\be
f=f(z), ~~~ f'=f'(z), ~~~ f''=f''(z), ~~~
s=s(z).
\ee
These quasiprimary operators transform as
\bea
&& T(z)=f'^2 T(f)+\f{c}{12}s, ~~~
   G(z)=f'^{3/2} G(f),  \nn\\
&& \mA(z)=f'^4\mA(f)+\f{5c+22}{30}s \lt( f'^2 T(f)+\f{c}{24}s \rt),\\
&& \mB(z)=f'^{7/2}\mB(f)+\f{4c+21}{48}f'^{3/2}sG(f), ~~~
   \mC(z)=f'^4\mC(f),                                                \nn\\
&& \mD(z)=f'^{9/2}\mD(f)+\f{\ii(10c-7)}{480}f'^{1/2}\lt( 3(2f''s-f's')G(f)+4f'^2sG'(f) \rt),\nn
\eea
from which transformations of their derivatives can be got easily.

\section{Two intervals on a line with zero temperature}\label{s3}

In this section we investigate the R\'enyi entropy of two short intervals on a line with zero temperature. In this case the CFT is located on a complex plane. We consider the case when the cross ratio $x$ is small, and so we can get the first few orders of the R\'enyi entropy in both the gravity side and the CFT side. In the gravity side it is the $\mN=1$ SUGRA, and there are contributions from both the graviton and the gravitino. In the CFT side, it is the $\mN=(1,1)$ SCFT, and there are contributions from both stress tensor $T$, $\bar T$ and operators $G$, $\bar G$.

\subsection{Holographic R\'enyi entropy}

The classical part of the holographic R\'enyi entropy is proportional to the central charge. It is related to the classical configuration of the gravity. The gravitino vanishes in classical SUGRA solution, and so we conclude that the gravitational configuration for pure Einstein gravity in \cite{Faulkner:2013yia,Barrella:2013wja} still applies to the $\mN=1$ SUGRA case. We have the classical part of the holographic R\'enyi mutual information \cite{Barrella:2013wja}
\bea \label{incl}
&& I_n^{(\cl)}=\frac{c (n-1) (n+1)^2 x^2}{144 n^3}+\frac{c (n-1) (n+1)^2 x^3}{144 n^3}
   +\frac{c (n-1) (n+1)^2 (11 n^2+1) (119 n^2-11) x^4}{207360 n^7}            \nn\\
&& \phantom{I_n^{(\cl)}=} +\frac{c (n-1) (n+1)^2 (589 n^4-2 n^2-11) x^5}{103680 n^7}+{O}(x^6),
\eea
with $c$ being the central charge of the dual SCFT.

The 1-loop part of the holographic R\'enyi entropy depends on the field contents of the gravity theory, and one considers the fluctuation of the fields around the classical background.
The procedure was given in \cite{Barrella:2013wja}, and it is related to the 1-loop partition function in \cite{Yin:2007gv,Giombi:2008vd}. The 1-loop R\'enyi entropy is
\be
S_n^{\oloop}=-\f{1}{n-1} \lt( \log Z_n^{\oloop} - n \log Z_1^{\oloop} \rt),
\ee
with $Z_n^{\oloop}$ being the 1-loop partition function around a genus $n-1$ handlebody background in the case of two intervals. When the spacetime is the quotient of global AdS$_3$ by a Schottky group $\G$, the 1-loop partition function for the spin-2 massless graviton is \cite{Yin:2007gv,Giombi:2008vd,Chen:2015uga}
\be \label{zoloop2}
Z^{\oloop}_{(2)}=\prod_{\g\in\mP}\prod_{m=0}^\inf \f{1}{|1-q_\g^{m+2}|},
\ee
with $\mP$ being a set of representatives of the primitive conjugacy classes of $\G$. Here $q_\g$ is defined in the way that the eigenvalues of $\g$ is $q_\g^{\pm1/2}$ with $|q_\g|<1$. For the case of two short intervals on a line with zero temperature, $q_\g$ can be written as expansion of the cross ration $x$, and so the 1-loop R\'enyi entropy can be expanded by $x$ too. To order $x^5$ the 1-loop R\'enyi mutual information is \cite{Barrella:2013wja}
\bea \label{inoloop2}
&& I_{n,(2)}^\oloop = \frac{(n+1) (n^2+11) (3 n^4+10 n^2+227) x^4}{3628800 n^7}                 \\
&& \phantom{I_{n,(2)}^\oloop =}
   +\frac{(n+1) (109 n^8+1495 n^6+11307 n^4+81905 n^2-8416) x^5}{59875200 n^9}+O(x^6).          \nn
\eea

In the $\mN=1$ SUGRA in AdS$_3$ background, there is also the superpartner of the graviton, the massless spin-3/2 gravitino. The 1-loop partition function (\ref{zoloop2}) should be multiplied by \cite{David:2009xg,Zhang:2012kya}
\be \label{zoloop32}
Z^{\oloop}_{(3/2)}=\prod_{\g\in\mP}\prod_{m=0}^\inf {|1+q_\g^{m+3/2}|}.
\ee
Then we get the additional 1-loop R\'enyi mutual information from the gravitino
\bea \label{inoloop32}
&& \hspace{-10mm} I_{n,(3/2)}^\oloop =  \frac{(n+1) (2 n^4+23 n^2+191) x^3}{60480 n^5}
                                      +\frac{(n+1) (33 n^6+358 n^4+2857 n^2-368) x^4}{604800 n^7}   \\
&& \hspace{-10mm} \phantom{I_{n,(3/2)}^\oloop =}
   +\frac{(n+1) (32422 n^8+336385 n^6+2606961 n^4-532285 n^2-24283) x^5}{479001600 n^9}+O(x^6).     \nn
\eea

\subsection{R\'enyi entropy in SCFT side}\label{s3.2}

We use the replica trick in the SCFT side, and get an SCFT on an $n$-sheeted complex plane, which is a genus $(n-1)(N-1)$ Riemann surface $\mR_{n,N}$ in the case of $N$ intervals. Equivalently, this configuration can be viewed as $n$ copies of the SCFT on a complex plane, with twist operators $\s$, $\td\s$ being inserted at the boundaries of the intervals \cite{Calabrese:2004eu}. We denote the $n$ copies of the SCFT as SCFT$^n$. The twist operators are primary operators with conformal weights
\be h_\s=\bar h_\s=h_{\td\s}=\bar h_{\td\s}=\f{c(n^2-1)}{24n}.\ee

We choose the two intervals $A=[0,y]\cup[1,1+y]$ with $y\ll1$, and so the cross ratio $x=y^2\ll1$. The partition of the SCFT on Riemann surface $\mR_{n,2}$ is equivalent of the four-point function of SCFT$^n$ on a complex plane \cite{Calabrese:2004eu}
\be \label{traran}
\tr_A\r_A^n=\lag \s(1+y,1+y)\td\s(1,1)\s(y,y)\td\s(0,0) \rag_C.
\ee
We use the OPE of the twist operators to do short interval expansion \cite{Headrick:2010zt,Calabrese:2010he,Chen:2013kpa,Chen:2013dxa,Perlmutter:2013paa,Chen:2014kja,Beccaria:2014lqa}. We denote the orthogonalized quasiprimary operators in SCFT$^n$ by $\Phi_K(z,\bar z)$, and a general $\Phi_K$ has normalization factor $\a_K$ and conformal weights $h_K$, $\bar h_K$.

In SCFT$^n$ we have the operator product expansion (OPE) \cite{Headrick:2010zt,Calabrese:2010he,Chen:2013kpa}
\be \label{ope}
\s(z,\bar z)\td \s(0,0)
=\f{c_n}{z^{2h_\s}\bar z^{2\bar h_\s}} \sum_K d_K \sum_{m,r\geq0} \f{a_K^m}{m!}\f{\bar a_K^r}{r!}
                                                                  z^{h_K+m}\bar z^{\bar h_K+r}
                                                                  \p^m \bar \p^r \Phi_K(0,0),
\ee
with $c_n$ being the normalization factor of the twist operators, summation $K$ being over all the independent quasiprimary operators of SCFT$^n$, and
\be
a_K^m\equiv \f{C_{h_K+m-1}^m}{C_{2h_K+m-1}^m}, ~~~ \bar a_K^r\equiv\f{C_{\bar h_K+r-1}^r}{C_{2\bar h_K+r-1}^r}.
\ee
Also, the OPE coefficient $d_K$ can be calculated as \cite{Calabrese:2010he}
\be
d_K=\f{1}{\a_K l^{h_K+\bar h_K}} \lim_{z\to\inf}z^{2 h_K}\bar z^{2\bar h_K}\lag \Phi_K(z,\bar z) \rag_{\mc R_{n,1}},
\ee
with $l$ being the length of the single interval $[0,l]$ that results in the Riemann surface $\mR_{n,1}$ in replica trick.
To calculate the expectation value of $\Phi_K$ on $\mR_{n,1}$, we use the conformal transformation \cite{Calabrese:2004eu,Calabrese:2010he}
\be
z \to f(z)=\Big( \f{z-l}{z} \Big)^{1/n},
\ee
that maps $\mR_{n,1}$ with coordinate $z$ to a complex plane with coordinate $f$.

With the OPE (\ref{ope}), the partition function (\ref{traran}) becomes \cite{Chen:2013kpa,Chen:2013dxa,Perlmutter:2013paa}
\be
\tr_A \r_A^n=c_n^2 x^{-\f{c(n^2-1)}{6n}}\sum_{K}\alpha_K d_K^2 x^{h_K+\bar{h}_K} F(h_K,h_K;2h_K;x) F(\bar h_K,\bar h_K;2\bar h_K;x),
\ee
with summation $K$ being over all the independent quasiprimary operators of SCFT$^n$, and $F$ being the hypergeometric function.
When every quasiprimary operator we consider can be written as a product of holomorphic and antiholomorphic parts and there is one-to-one correspondence between operators in holomorphic and antiholomorphic sectors, the partition function can be further simplified as
\be
\tr_A \r_A^n=c_n^2 x^{-\f{c(n^2-1)}{6n}} \Big( \sum_{K}\alpha_K d_K^2 x^{h_K} F(h_K,h_K;2h_K;x) \Big)^2,
\ee
with summation $K$ being over all the independent holomorphic quasiprimary operators. In this case the R\'enyi mutual information is
\be \label{in}
I_n=\f{2}{n-1}\log \Big( \sum_{K} \alpha_K d_K^2 x^{h_K} F(h_K,h_K;2h_K;x) \Big).
\ee

\begin{table}
  \centering
\begin{tabular}{|c|c|c|c|}\hline
  level & quasiprimary operator & degeneracy & number \\\hline

  0 & 1 & 1 & 1 \\ \hline

  3/2 & $G_j$ & $n$ & $n$ \\\hline

   2 & $T_j$ & $n$ & $n$ \\\hline

   3 & $G_{j_1}G_{j_2}$ with $j_1<j_2$ & $\f{n(n-1)}{2}$ & $\f{n(n-1)}{2}$ \\\hline

  \multirow{2}*{7/2} & $\mc B_j$                           & $n$      & \multirow{2}*{$n^2$} \\ \cline{2-3}
                     &$T_{j_1}G_{j_2}$ with $j_1 \neq j_2$ & $n(n-1)$ &                      \\\hline

                   & $\mc A_j$                       & $n$             &                         \\ \cline{2-3}
  \multirow{2}*{4} & $\mc C_j$                       & $n$             & \multirow{2}*{$n(n+1)$} \\ \cline{2-3}
                   & $T_{j_1}T_{j_2}$ with $j_1<j_2$ & $\f{n(n-1)}{2}$ &                         \\ \cline{2-3}
                   & $\mc E_{j_1j_2}$ with $j_1<j_2$ & $\f{n(n-1)}{2}$ &                         \\ \hline

      & $\mD_j$                                    & $n$                  &                      \\ \cline{2-3}
  9/2 & $G_{j_1}G_{j_2}G_{j_3}$ with $j_1<j_2<j_3$ & $\f{n(n-1)(n-2)}{6}$ & $\f{n(n+1)(n+2)}{6}$ \\ \cline{2-3}
      & $\mF_{j_1j_2}$ with $j_1 \neq j_2$         & $n(n-1)$             &                      \\ \hline

                   & $G_{j_1}\mc B_{j_2}$ with $j_1 \neq j_2$                              & $n(n-1)$             &                                     \\ \cline{2-3}
  \multirow{2}*{5} & $T_{j_1}G_{j_2}G_{j_3}$ with $j_1\neq j_2$, $j_1\neq j_3$, $j_2< j_3$ & $\f{n(n-1)(n-2)}{2}$ & \multirow{2}*{$\f{n(n-1)(n+2)}{2}$} \\ \cline{2-3}
                   & $\mH_{j_1j_2}$ with $j_1<j_2$                                         & $\f{n(n-1)}{2}$      &                                     \\ \cline{2-3}
                   & $\mI_{j_1j_2}$ with $j_1<j_2$                                         & $\f{n(n-1)}{2}$      &                                     \\ \hline

    $\cdots$ & $\cdots$ & $\cdots$ & $\cdots$\\
    \hline
\end{tabular}
\caption{Holomorphic quasiprimary operators in SCFT$^n$ to level 5. Here $j$, $j_1$, $j_2$, $j_3$ are integers and take values from $0$ to $n-1$. }
\label{PhiK}
\end{table}

In SCFT$^n$, we count the number of independent holomorphic quasiprimary operators as
\bea
&& (1-x) \lt(\tr x^{L_0}\rt)^n + x = 1+n x^{3/2}+n x^2+\frac{n(n-1)}{2} x^3+n^2 x^{7/2}+n (n+1) x^4 \\
&& \ph{1-x) \lt(\tr x^{L_0}\rt)^n + x =}
   +\frac{n (n+1) (n+2)}{6} x^{9/2}+\frac{n (n-1)(n+2)}{2} x^5+O(x^{11/2}),                          \nn
\eea
with $\tr x^{L_0}$ being defined as (\ref{trxl0}). These holomorphic quasiprimary operators are listed in Table~\ref{PhiK}, where we have the definitions
\bea
&& \mE_{j_1j_2}=G_{j_1}\ii\p G_{j_2} - \ii\p G_{j_1}G_{j_2},  ~~~
   \mF_{j_1j_2}=T_{j_1}\ii\p G_{j_2} - \f{3}{4} \ii\p T_{j_1}G_{j_2},\\
&& \mH_{j_1j_2}=T_{j_1}\ii\p T_{j_2} - \ii\p T_{j_1}T_{j_2},  ~~~
   \mI_{j_1j_2}=\p G_{j_1}\p G_{j_2} - \f{3}{8} ( G_{j_1}\p^2 G_{j_2}+\p^2 G_{j_1}G_{j_2} ).\nn
\eea
The factors $\ii$'s in $\mF_{j_1j_2}$ and $\mH_{j_1j_2}$ are chosen to make $\a_\mF>0$ and $\a_\mH>0$. We would have $\a_\mE>0$ and $\a_\mI>0$ \emph{if} $G$ is bosonic. But in fact $G$ is fermionic, and so we have $\a_\mE<0$ and $\a_\mI<0$ in our definitions.

Then we calculate the normalization factors $\a_K$ and OPE coefficients $d_K$.
It is easy to see that
\be \label{dK0}
d_G=d_\mB=d_\mC=d_\mD=d_{TG}=d_{GGG}=d_\mF=0.
\ee
The useful normalization factors are
\bea \label{alphaK}
&& \a_1=1, ~~~
   \a_T=\f{c}{2}, ~~~
   \a_{GG}=-\f{4c^2}{9}, ~~~
   \a_\mA=\f{c(5c+22)}{10},            \nn\\
&& \a_{TT}=\f{c^2}{4}, ~~~
   \a_\mE=-\f{8c^2}{3}, ~~~
   \a_{G\mB}=-\f{c^2(4c+21)}{18},      \\
&& \a_{TGG}=-\f{2c^3}{9}, ~~~
   \a_\mH=2c^2, ~~~
   \a_\mI=-7c^2.                      \nn
\eea
The useful OPE coefficients are
\bea \label{dK}
&& d_1=1, ~~~
   d_T=\f{n^2-1}{12n^2}, ~~~
   d_{GG}^{j_1j_2}=-\f{3\ii}{16n^3c}\f{1}{s_{j_1j_2}^3},                                                                            \nn\\
&& d_\mA=\f{(n^2-1)^2}{288n^4}, ~~~
   d_{TT}^{j_1j_2}=\f{1}{8n^4c}\f{1}{s_{j_1j_2}^4}+\f{(n^2-1)^2}{144n^4},                                                           \nn\\
&& d_\mE^{j_1j_2}=-\f{3\ii}{32n^4c}\f{c_{j_1j_2}}{s_{j_1j_2}^4}, ~~~
   d_{G\mB}^{j_1j_2}=-\f{\ii(n^2-1)}{64n^5c}\f{1}{s_{j_1j_2}^3},                                                                    \\
&& d_{TGG}^{j_1j_2j_3}=\f{\ii}{64n^5c^2}\bigg( \f{9}{s_{j_1j_2}^2 s_{j_1j_3}^2 s_{j_2j_3}}-\f{(n^2-1)c}{s_{j_2j_3}^3} \bigg),         \nn\\
&& d_\mH^{j_1j_2}=\f{1}{16n^5c}\f{c_{j_1j_2}}{s_{j_1j_2}^5}, ~~~
   d_\mI^{j_1j_2}=-\f{\ii}{448n^5c}\bigg( \f{28}{s_{j_1j_2}^5}-\f{3(n^2+7)}{s_{j_1j_2}^3} \bigg),                                     \nn
\eea
with the definitions  $s_{j_1j_2}\equiv\sin\f{\pi(j_1-j_2)}{n}$ and $c_{j_1j_2}\equiv\cos\f{\pi(j_1-j_2)}{n}$.

Finally, using the formula (\ref{in}), normalization factors (\ref{alphaK}), OPE coefficients (\ref{dK0}) and (\ref{dK}), as well as the summation formulas in Appendix~\ref{sa}, we obtain the R\'enyi mutual information
\be
I_n=I_n^{\tree}+I_n^{\oloop}+I_n^{\tloop}+\cdots,
\ee
with the tree part being
\bea \label{intree}
&&\hspace{-10mm} I_n^{\tree} = \frac{c (n-1) (n+1)^2 x^2}{144 n^3}+\frac{c (n-1) (n+1)^2 x^3}{144 n^3}
                 +\frac{c (n-1) (n+1)^2 (11 n^2+1) (119 n^2-11) x^4}{207360 n^7}         \nn\\
&&\hspace{-10mm} \phantom{I_n^{\tree}=}
                 +\frac{c (n-1) (n+1)^2 (589 n^4-2 n^2-11) x^5}{103680 n^7}+O(x^6),
\eea
the 1-loop part being
\bea \label{inoloop}
&& I_n^{\oloop}=\frac{(n+1) (2 n^4+23 n^2+191) x^3}{60480 n^5}+\frac{(n+1) (201 n^6+2191 n^4+17479 n^2+289) x^4}{3628800 n^7}  \nn\\
&& \phantom{I_n^{\oloop}=}
                +\frac{(n+1) (11098 n^8+116115 n^6+899139 n^4+40985 n^2-30537) x^5}{159667200 n^9}+O(x^6),
\eea
and the 2-loop part being
\be
I_n^{\tloop}=\frac{(n+1)(n^2-4)(n^2+19) (n^4+19 n^2+628) x^5}{26611200 c n^9}+O(x^6).
\ee

The result in the SCFT side can be compared to the SUGRA one. The tree part of the R\'enyi mutual information (\ref{intree}) equals the classical part of the holographic R\'enyi mutual information (\ref{incl})
\be
 I_n^{\tree}= I_n^{\cl}.
\ee
The 1-loop part of the R\'enyi mutual information (\ref{inoloop}) equals the summation of the 1-loop holographic R\'enyi mutual information from the graviton (\ref{inoloop2}) and gravitino (\ref{inoloop32})
\be
 I_n^{\oloop}= I_{n,(2)}^{\oloop}+I_{n,(3/2)}^{\oloop}.
\ee
The result is in accordance with the SUGRA/SCFT correspondence.

\section{One interval on a circle with low temperature}\label{s4}

In this section we investigate the R\'enyi entropy of one interval on a circle with low temperature. In this case the CFT is located on a torus.
We calculate in both the SUGRA and SCFT sides, using the methods in \cite{Barrella:2013wja,Cardy:2014jwa,Chen:2014unl,Chen:2015uia}.

\subsection{Holographic R\'enyi entropy}

We set that the length of the circle is $R$ and the interval is $A=[-l/2,l/2]$. The temperature is $T$, and the inverse temperature is $\b=1/T$. In low temperature we have $\b \gg R$, and the holographic R\'enyi entropy can be expanded by $\exp(-2\pi\b/R)$ \cite{Barrella:2013wja,Chen:2014unl,Chen:2015uia}. The procedure is that one firstly calculates the R\'enyi entropy at large temperature that is expanded by $\exp(-2\pi R/\b)$ and then makes the modular transformation $R \to \ii\b$, $\b\to \ii R$ to get the R\'enyi entropy at low temperature.

At zero temperature, the holographic R\'enyi entropy for one interval with length $l$ on a circle with length $R$ is \cite{Ryu:2006bv,Ryu:2006ef,Barrella:2013wja}
\be
S_n=\f{c(n+1)}{6n}\log \Big( \f{R}{\pi\e}\sin\f{\pi l}{R} \Big),
\ee
with $\e$ being the UV cutoff. This is the same as the CFT result (\ref{e12}) in \cite{Calabrese:2004eu}.
At low temperature, there would be thermal correction to the R\'enyi entropy.
Similar to the case of two intervals on a line, the classical Holographic R\'enyi entropy in SUGRA is the same as that in pure Einstein gravity.
One can find the classical part of the correction to the holographic R\'enyi entropy in \cite{Barrella:2013wja,Chen:2014unl,Chen:2015uia}
\be \label{dsncl}
\d S_n^\cl=- \lt( \f{c(n-1)(n+1)^2}{9n^3}\sin^4\f{\pi l}{R} \rt)\ep^{-4\pi\b/R}+O(\ep^{-6\pi\b/R}).
\ee
For the 1-loop part, (\ref{zoloop2}) and (\ref{zoloop32}) still apply, but now the Schottky group is parameterized differently. The 1-loop correction to R\'enyi entropy from graviton can be found in \cite{Barrella:2013wja,Chen:2014unl,Chen:2015uia}
\be \label{dsnoloop2}
\d S^\oloop_{n,(2)}=-\f{1}{n-1}\lt[ \lt( \f{2}{n^3}\f{\sin^4\f{\pi l}{R}}{\sin^4\f{\pi l}{nR}}-2n \rt)\ep^{-4\pi\b/R}+O(\ep^{-6\pi\b/R})  \rt].
\ee
We also get the 1-loop correction to the R\'enyi entropy from gravitino
\bea \label{dsnoloop32}
&&\hspace{-10mm} \d S^\oloop_{n,(3/2)}=-\f{1}{n-1}\bigg\{ \bigg( \f{2}{n^2}\f{\sin^3\f{\pi l}{R}}{\sin^3\f{\pi l}{nR}}-2n \bigg)\ep^{-3\pi\b/R}
   + \bigg[ \f{1}{n^4}\f{\sin^3\f{\pi l}{R}}{\sin^5\f{\pi l}{nR}}\bigg( 6n^2\cos^2\f{\pi l}{R}\sin^2\f{\pi l}{nR}      \\
&&\hspace{-10mm} \phantom{\d S^\oloop_{n,(3/2)}=}
                                                               -3n\sin\f{2\pi l}{R}\sin\f{2\pi l}{nR}
                                                               +\sin^2\f{\pi l}{R}\Big( 3\cos\f{2\pi l}{nR}+5 \Big) \bigg) -2n \bigg] \ep^{-5\pi\b/R}
   +O(\ep^{-6\pi\b/R})
             \bigg\}.                   \nn
\eea
We take the $n\to 1$ limit, and get the 1-loop correction to the entanglement entropy
\bea
&& \d S^\oloop_{(2)}=8 \Big( 1-\f{\pi l}{R}\cot\f{\pi l}{R} \Big)\ep^{-4\pi\b/R}+O(\ep^{-6\pi\b/R}),      \\
&& \d S^\oloop_{(3/2)}= 6 \Big( 1-\f{\pi l}{R}\cot\f{\pi l}{R} \Big)\ep^{-3\pi\b/R}
                       +10 \Big( 1-\f{\pi l}{R}\cot\f{\pi l}{R} \Big)\ep^{-5\pi\b/R}
                       +O(\ep^{-6\pi\b/R}).                                                               \nn
\eea

\subsection{R\'enyi entropy in SCFT side}

We use the method in \cite{Cardy:2014jwa,Chen:2014unl,Chen:2015uia} and calculate the contributions of stress tensor $T$, $\bar T$ and operators $G$, $\bar G$ to R\'enyi entropy in the SCFT side. As in the case of two intervals, we only analyze the holomorphic operators carefully, and we multiply the R\'enyi entropy by a factor 2 to account for the contributions from the antiholomorphic sector.

When the temperature is low $\b\gg R$, we has the SCFT that is located on a cylinder with a thermally corrected density matrix.
The hamiltonian from the holomorphic sector is
\be
H=\f{2\pi}{R} \Big( L_0-\f{c}{24} \Big),
\ee
and without affecting the final result we shift it to be
\be
H=\f{2\pi}{R} L_0.
\ee
We have the unnormalized density matrix \cite{Cardy:2014jwa,Chen:2014unl,Chen:2015uia}
\be
\r=\uum\vac+\sum_\phi\sum_{m=0}^\inf \f{\ep^{-2\pi(m+h_\phi)\b/R}}{\a_{\p^m\phi}} |\p^m\phi\rag \lag\p^m\phi|,
\ee
with summation $\phi$ being over all the independent non-identity holomorphic quasiprimary operators of the SCFT. To the order of $\ep^{-5\pi\b/R}$ we only need to consider three states $|G\rag$, $|T\rag$, and $|\p G\rag$.

We trace the degree of freedom of $B$, and get the reduced density matrix $\r_A=\tr_B \r$. Then we get the R\'enyi entropy
\be
S_n=-\f{2}{n-1} \log \f{\tr_A \r_A^n}{(\tr_A \r_A)^n},
\ee
with the additional factor 2 accounting for contributions from the antiholomorphic sector.
Note that we have $\tr_A\r_A=\tr\r$, as well as \cite{Calabrese:2004eu}\footnote{Note that here we have only incorporated contributions from holomorphic sector, and to get the full result we need to multiply a factor 2 on the right-hand side of the equation.}
\be
\log \tr_A (\tr_B\uum\vac)^n=-\f{c(n^2-1)}{12n}\log \Big( \f{R}{\pi\e}\sin\f{\pi l}{R} \Big).
\ee
Thus the thermal correction to R\'enyi entropy is
\be
\d S_n=-\f{2n}{n-1}\big[ (I-1)\ep^{-3\pi\b/R} +(I\!I-1)\ep^{-4\pi\b/R} +(I\!I\!I-1)\ep^{-5\pi\b/R} +O(\ep^{-6\pi\b/R}) \big],
\ee
with definitions of $I$, $I\!I$ and $I\!I\!I$ being
\bea \label{IIIIII}
&& I=\f{\tr_A \big[ \tr_B |G\rag\lag G| (\tr_B \uum\vac)^{n-1} \big]}{\a_G\tr_A\big(\tr_B \uum\vac\big)^n},  \nn\\
&& I\!I=\f{\tr_A \big[ \tr_B |T\rag\lag T| (\tr_B \uum\vac)^{n-1} \big]}{\a_T\tr_A\big(\tr_B \uum\vac\big)^n},  \\
&& I\!I\!I=\f{\tr_A \big[ \tr_B |{\p G}\rag\lag {\p G}| (\tr_B \uum\vac)^{n-1} \big]}{\a_{\p G}\tr_A\big(\tr_B \uum\vac\big)^n}.  \nn
\eea

Originally we have the SCFT on a cylinder with coordinate $w=x-\ii t$ and  of circumference $R$. We denote the cylinder also by $R$. In replica trick we get an SCFT on an $n$-sheeted cylinder, which we denote by $R^n$. Note that here we take the viewpoint that there is one copy of the SCFT and $n$ copies of the cylinder. Firstly we make the transformation
\be
z=\ep^{2\pi\ii w/R},
\ee
and this changes the $n$-sheeted cylinder $R^n$ with coordinate $w$ to an $n$-sheeted complex plane $C^n$ with coordinate $z$. Then we make the transformation \cite{Cardy:2014jwa,Chen:2014unl,Chen:2015uia}
\be \label{fz}
f(z) = \bigg( \f{z-\ep^{\ii\pi l/R}}{z-\ep^{-\ii\pi l/R}} \bigg)^{1/n},
\ee
and this changes the $n$-sheeted complex plane $C^n$ to a complex plane $C$ with coordinate $f$.
To calculate (\ref{IIIIII}), we adopt the strategy in \cite{Chen:2014unl,Chen:2015uia}. Firstly they equal to correlation functions on $C^n$, and then one uses (\ref{pmphiinf}), (\ref{fz}) and transforms them to correlation functions on $C$. Explicitly, we have
\bea
&& I=\f{\lag G(\inf)G(0)\rag_{C^n}}{\a_G}= \f{1}{n^3}\f{\sin^3\f{\pi l}{R}}{\sin^3\f{\pi l}{nR}},  \nn\\
&& I\!I=\f{\lag T(\inf)T(0)\rag_{C^n}}{\a_T}=  \f{c(n^2-1)^2}{18n^4} \sin^4\f{\pi l}{R}
                                              +\f{1}{n^4}\f{\sin^4\f{\pi l}{R}}{\sin^4\f{\pi l}{nR}},  \\
&& I\!I\!I=\f{\lag \p G(\inf)\p G(0)\rag_{C^n}}{\a_{\p G}}
          =\f{1}{2n^5}\f{\sin^3\f{\pi l}{R}}{\sin^5\f{\pi l}{nR}}\bigg( 6n^2\cos^2\f{\pi l}{R}\sin^2\f{\pi l}{nR}
                                                                       -3n\sin\f{2\pi l}{R}\sin\f{2\pi l}{nR}     \nn\\
&&\hspace{6.8cm}
                                                                       +\sin^2\f{\pi l}{R}\Big( 3\cos\f{2\pi l}{nR}+5 \Big) \bigg).  \nn
\eea

Then we get the tree part of the correction to the R\'enyi entropy
\be
\d S_n^\tree=- \lt( \f{c(n-1)(n+1)^2}{9n^3}\sin^4\f{\pi l}{R} \rt)\ep^{-4\pi\b/R}+O(\ep^{-6\pi\b/R}),
\ee
which equals to the classical part of the correction to holographic R\'enyi entropy $\d S_n^\cl$ (\ref{dsncl}). The 1-loop part of correction to R\'enyi entropy is
\bea
&&\hspace{-10mm} \d S^\oloop_{n}=-\f{1}{n-1}\bigg\{ \bigg( \f{2}{n^2}\f{\sin^3\f{\pi l}{R}}{\sin^3\f{\pi l}{nR}}-2n \bigg)\ep^{-3\pi\b/R}
   + \lt( \f{2}{n^3}\f{\sin^4\f{\pi l}{R}}{\sin^4\f{\pi l}{nR}}-2n \rt)\ep^{-4\pi\b/R}             \nn\\
&&\hspace{2cm} + \bigg[ \f{1}{n^4}\f{\sin^3\f{\pi l}{R}}{\sin^5\f{\pi l}{nR}}\bigg( 6n^2\cos^2\f{\pi l}{R}\sin^2\f{\pi l}{nR}
                                                                          -3n\sin\f{2\pi l}{R}\sin\f{2\pi l}{nR}      \\
&&\hspace{2.5cm} +\sin^2\f{\pi l}{R}\Big( 3\cos\f{2\pi l}{nR}+5 \Big) \bigg) -2n \bigg] \ep^{-5\pi\b/R}
                 +O(\ep^{-6\pi\b/R})  \bigg\},  \nn
\eea
and this equals the summation of contributions of graviton and gravitino to the 1-loop holographic R\'enyi entropy $\d S^\oloop_{n,(2)}$  (\ref{dsnoloop2}) and $\d S^\oloop_{n,(3/2)}$  (\ref{dsnoloop32}).

\section{Conclusion and discussion}\label{s5}

In this paper we investigated the holographic R\'enyi entropy for the two-dimensional $\mN=(1,1)$ SCFT, which is dual to $\mN=1$ SUGRA in AdS$_3$ spacetime.
We considered both cases of two short intervals on a line with zero temperature and one interval on a circle with low temperature.
For the first case, we got the R\'enyi mutual information to order $x^5$ with $x$ being the cross ratio.
For the second case, we got the thermal correction of R\'enyi entropy to order $\ep^{-5\pi\b/R}$ with $\b$ being the inverse temperature and $R$ being the length of the circle.
We found perfect matches between SUGRA and SCFT results.

It would be nice to extend the results to higher orders, in terms of $x$ for the two intervals case and in terms of $\ep^{-\pi\b/R}$ for the one interval case. In the SUGRA side, the so-called $p$-consecutively decreasing words and $p$-letter words of Schottky group with $p\geq2$ would be needed. In SCFT side one needs $m$-point correlation functions with $m\geq4$. It is also interesting to consider the holographic R\'enyi entropy of large interval at high temperature in the SUGRA/SCFT correspondence, as what was done for Einstein gravity in \cite{Chen:2014hta,Chen:2015kua}.

In this paper we have only considered the NS sector of the SCFT. It is an interesting question whether one can calculate R\'enyi entropy of the SCFT in R sector vacuum and compare it with the holographic result in some suitable gravitational background.

\section*{Acknowledgments}

The author would like to thank Bin Chen and Jun-Bao Wu for careful reading of the manuscript and valuable suggestions.
Special thanks Matthew Headrick for his Mathematica code \emph{Virasoro.nb} that could be downloaded at \url{http://people.brandeis.edu/~headrick/Mathematica/index.html}.
The work was in part supported by NSFC Grants No.~11222549 and No.~11575202.

\appendix

\section{Some useful summation formulas}\label{sa}

In the appendix we give some summation formulas that are used in our calculation. We define
\be
f_m=\sum_{j=1}^{n-1}\f{1}{ \lt( \sin\f{\pi j}{n} \rt)^{2m}},
\ee
and explicitly we need
\bea
&& f_1=\frac{n^2-1}{3}, ~~~ f_2=\frac{(n^2-1) \left(n^2+11\right)}{45} , \nn\\
&& f_3=\frac{(n^2-1)  \left(2 n^4+23 n^2+191\right)}{945} ,    \\
&& f_4=\frac{(n^2-1) \left(n^2+11\right) \left(3 n^4+10 n^2+227\right)}{14175},  \nn\\
&& f_5=\frac{(n^2-1) \left(2 n^8+35 n^6+321 n^4+2125 n^2+14797\right)}{93555}.\nn
\eea
The above formulas are useful because they appear in the following summations
\be
\sum_{0\leq j_1 <j_2 \leq n-1} \f{1}{s^{2m}_{j_1j_2}}=\f{n}{2}f_m,  ~~~
\sum_{0\leq j_1 <j_2<j_3 \leq n-1} \lt( \f{1}{s^{2m}_{j_1j_2}} +\f{1}{s^{2m}_{j_2j_3}} +\f{1}{s^{2m}_{j_3j_1}} \rt)=\f{n(n-2)}{2}f_m,
\ee
with $s_{j_1j_2}\equiv\sin\f{\pi(j_1-j_2)}{n}$.
There are also two other useful summation formulas
\bea
&&\hspace{-10mm} \sum_{0\leq j_1 <j_2<j_3 \leq n-1}
   \f{1}{s^2_{j_1j_2}s^2_{j_2j_3}s^2_{j_3j_1}}
   \lt( \f{1}{s^2_{j_1j_2}}+\f{1}{s^2_{j_2j_3}}+\f{1}{s^2_{j_3j_1}} \rt)
   =
   \frac{n (n^2-1) (n^2-4) (n^4+40 n^2+679)}{14175},  \nn\\
&&\hspace{-10mm} \sum_{0\leq j_1 <j_2<j_3 \leq n-1}
    \f{s^2_{j_1j_2}+s^2_{j_2j_3}+s^2_{j_3j_1}}{s^4_{j_1j_2}s^4_{j_2j_3}s^4_{j_3j_1}}
    =
    \frac{2 n (n^2-1) (n^2-4) (n^2+19) (n^4+19 n^2+628)}{467775}.
\eea


\providecommand{\href}[2]{#2}\begingroup\raggedright\endgroup

\end{document}